\documentclass[prd,twocolumn,twoside,preprintnumbers,superscriptaddress,nofootinbib]{revtex4}
\usepackage{amsmath,slashed}
\usepackage{graphicx,graphics,color}
\usepackage{dcolumn}
\usepackage[hyperfootnotes=false]{hyperref}
\usepackage{xspace}
\usepackage{enumerate}
\usepackage{varwidth}

\newcommand{\MeV}{\,\text{MeV}}
\newcommand{\GeV}{\,\text{GeV}}
\newcommand{\TeV}{\,\text{TeV}}

\newcommand{\beq}{\begin{equation}}
\newcommand{\eeq}{\end{equation}}
\renewcommand{\Re}{\text{Re}\,}

\allowdisplaybreaks[1]

\begin{document}

\preprint{CERN-TH-2021-017, PSI-PR-21-02, ZU-TH 04/21}
\title{
The Fermi constant from muon decay versus electroweak fits and CKM unitarity}

\author{Andreas Crivellin}
\affiliation{CERN Theory Division, CH--1211 Geneva 23, Switzerland}
\affiliation{Physik-Institut, Universit\"at Z\"urich, Winterthurerstrasse 190, CH--8057 Z\"urich, Switzerland}
\affiliation{Paul Scherrer Institut, CH--5232 Villigen PSI, Switzerland}
\author{Martin Hoferichter}
\affiliation{Albert Einstein Center for Fundamental Physics, Institute for Theoretical Physics, University of Bern, Sidlerstrasse 5, CH--3012 Bern, Switzerland}
\author{Claudio Andrea Manzari}
\affiliation{Physik-Institut, Universit\"at Z\"urich, Winterthurerstrasse 190, CH--8057 Z\"urich, Switzerland}
\affiliation{Paul Scherrer Institut, CH--5232 Villigen PSI, Switzerland}

\begin{abstract}
The Fermi constant ($G_F$) is extremely well measured through the muon lifetime, defining one of the key fundamental parameters in the Standard Model (SM). Therefore, to search for physics beyond the SM (BSM) via $G_F$, the constraining power is determined by the precision of the second-best independent determination of $G_F$. The best alternative extractions of $G_F$ proceed either via the global electroweak (EW) fit or from superallowed $\beta$ decays in combination with the Cabibbo angle measured in kaon, $\tau$, or $D$ decays. Both variants display some tension with $G_F$ from muon decay, albeit in opposite directions, reflecting the known tensions within the EW fit and hints for the apparent violation of CKM unitarity, respectively. We investigate how BSM physics could bring the three determinations of $G_F$ into agreement using SM effective field theory and comment on future perspectives.       
\end{abstract}

\maketitle

\section{Introduction}
 
The numerical value of the Fermi constant $G_F$ is conventionally defined via the muon lifetime within the SM. Even though this measurement is extremely precise~\cite{Webber:2010zf,Tishchenko:2012ie,Gorringe:2015cma}
\begin{equation}
\label{Mulan}
G_F^\mu=1.1663787(6) \times 10^{-5} \GeV^{-2},
\end{equation}
at the level of $0.5\,\text{ppm}$, its determination of the Fermi constant is not necessarily free of BSM contributions. In fact, one can only conclude the presence or absence of BSM effects by comparing $G_F^\mu$ to another independent determination. This idea was first introduced by Marciano in Ref.~\cite{Marciano:1999ih}, concentrating on $Z$-pole observables and the fine-structure constant $\alpha$. In addition to a lot of new data that have become available since 1999, another option already mentioned in Refs.~\cite{Marciano:1999ih,Lee:1977tib}---the determination of $G_F$ from $\beta$ and kaon decays using CKM unitarity---has become of particular interest due to recent hints for the (apparent) violation of first-row CKM unitarity. These developments motivate a fresh look at the Fermi constant, in particular on its extraction from a global EW fit and via CKM unitarity, as will be discussed in the first part of this Letter.   

The comparison of the resulting values for $G_F$ then serves as a model-independent measure of possible BSM effects. It shows that with modern input the two independent extractions are close in precision, yet still lagging behind muon decay by almost three orders of magnitude. Therefore, the BSM sensitivity is governed by the uncertainty of these two indirect determinations. Since the different $G_F$ measurements turn out to display some disagreement beyond their quoted uncertainties, the second part of this Letter is devoted to a systematic analysis of possible BSM contributions in SM effective field theory (SMEFT)~\cite{Buchmuller:1985jz,Grzadkowski:2010es} to see which scenarios could account for these tensions without being excluded by other constraints. This is important to identify BSM scenarios that could be responsible for the tensions, which 
will be scrutinized with forthcoming data in the next years.

\section{Determinations of $\boldsymbol{G_F}$}

Within the SM, the Fermi constant $G_F$ is defined by, and is most precisely determined from, the muon lifetime~\cite{Tishchenko:2012ie}
\begin{equation}
\frac{1}{\tau_{\mu}}=\frac{(G_F^\mu)^2m_{\mu}^5}{192\pi^3}(1+\Delta q),
\end{equation}
where $\Delta q$ includes the phase space, QED, and hadronic radiative corrections. The resulting numerical value in Eq.~\eqref{Mulan} is so precise that its error can be ignored in the following. To address the question whether $G_F^\mu$ subsumes BSM contributions, however, alternative independent determinations of $G_F$ are indispensable, and their precision limits the extent to which BSM contamination in $G_F^\mu$ can be detected.  

In Ref.~\cite{Marciano:1999ih}, the two best independent determinations were found as
\begin{align}
\label{Marciano}
 G_F^{Z\ell^+\ell^-}&=1.1650(14)\big({}^{+11}_{-6}\big)\times 10^{-5}\GeV^{-2},\notag\\
 G_F^{(3)}&=1.1672(8)\big({}^{+18}_{-7}\big)\times 10^{-5}\GeV^{-2},
\end{align}
where the first variant uses the width for $Z\to\ell^+\ell^-(\gamma)$, while the second employs $\alpha$ and $\sin^2\theta_W$, together with the appropriate radiative corrections. Since the present uncertainty in $\Gamma[Z\to\ell^+\ell^-(\gamma)]=83.984(86)\MeV$~\cite{Zyla:2020zbs} is only marginally improved compared to the one available in 1999, the update
\begin{equation}
 G_F^{Z\ell^+\ell^-}=1.1661(16)\times 10^{-5}\GeV^{-2}
\end{equation}
does not lead to a gain in precision, but the shift in the central value improves agreement with $G_F^\mu$. The second variant, $G_F^{(3)}$, is more interesting, as here the main limitation arose from the radiative corrections, which have seen significant improvements regarding the input values for the masses of the top quark, $m_t$, the Higgs boson, $M_H$, the strong coupling, $\alpha_s$, and the hadronic running of $\alpha$. In fact, with all EW parameters determined, it now makes sense to use the global EW fit, for which $G_F^\mu$ is usually a key input quantity, instead as a tool to determine $G_F$ in a completely independent way. 

\begin{figure}[t]
	\includegraphics[width=\linewidth]{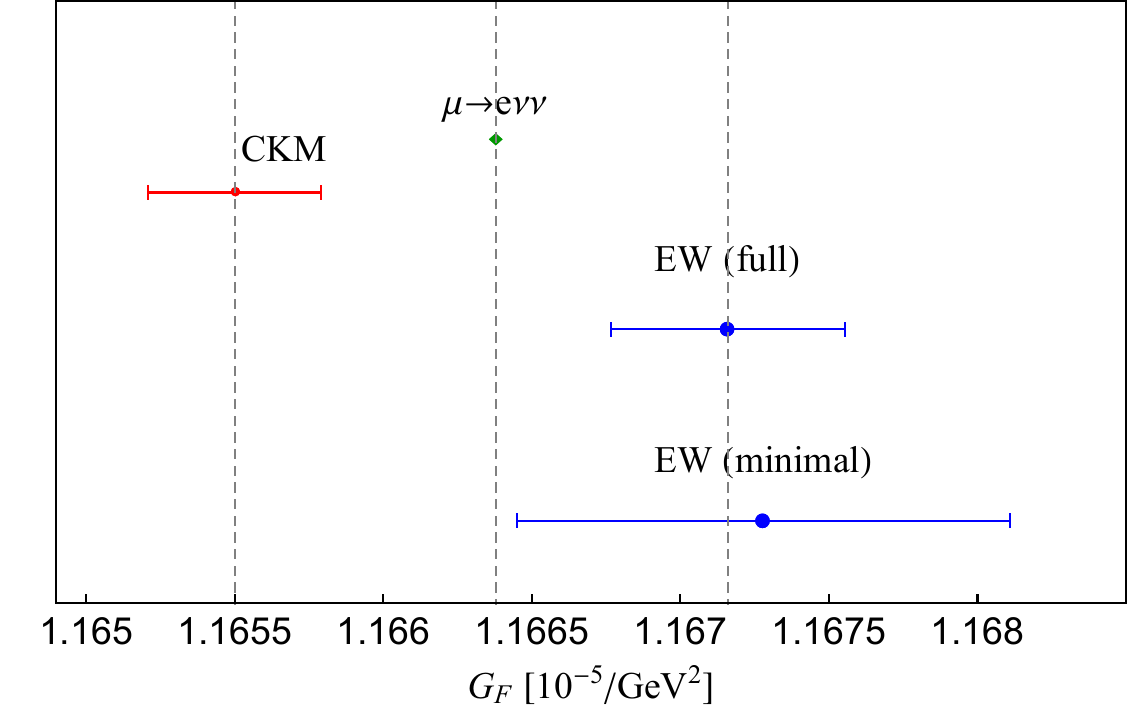}
	\caption{Values of $G_F$ from the different determinations.\label{GFplot} }
\end{figure}

The EW observables included in our fit ($W$ mass, $\sin^2\theta_W$, and  
$Z$-pole observables~\cite{Schael:2013ita,ALEPH:2005ab}) are given in Table~\ref{ObsEW}, with the other input parameters  summarized in Table~\ref{ParamEW}. Here, the hadronic running $\Delta\alpha_\text{had}$ is taken from $e^+e^-$ data, which are insensitive to the changes  in $e^+e^-\to\text{hadrons}$ cross sections~\cite{Aoyama:2020ynm,Davier:2017zfy,Keshavarzi:2018mgv,Colangelo:2018mtw,Hoferichter:2019mqg,Davier:2019can,Keshavarzi:2019abf,Hoid:2020xjs} recently suggested by lattice QCD~\cite{Borsanyi:2020mff}, as long as these changes are concentrated at low energies~\cite{Crivellin:2020zul,Keshavarzi:2020bfy,Malaescu:2020zuc,Colangelo:2020lcg}.  
We perform the global EW fit
in a Bayesian framework using the publicly available \texttt{HEPfit} package~\cite{deBlas:2019okz}, whose Markov Chain Monte Carlo (MCMC) determination of posteriors is powered by the Bayesian Analysis Toolkit (\texttt{BAT})~\cite{Caldwell:2008fw}. As a result, we find 
\begin{equation}
\label{GFEW}
G_F^\text{EW}\Big|_\text{full}=1.16716(39)\times 10^{-5}\GeV^{-2},
\end{equation}
a gain in precision over $G_F^{(3)}$ in Eq.~\eqref{Marciano} by a factor $5$. As depicted in Fig.~\ref{GFplot}, this value lies above $G_F^\mu$ by $2\sigma$, reflecting the known tensions within the EW fit~\cite{Baak:2014ora,deBlas:2016ojx}. For comparison, we also considered a closer analog of $G_F^{(3)}$, by only keeping $\sin^2\theta_W$ from Table~\ref{ObsEW} in the fit, which gives 
\begin{equation}
G_F^\text{EW}\Big|_\text{minimal}=1.16728(83)\times 10^{-5}\GeV^{-2},
\end{equation}
consistent with Eq.~\eqref{GFEW}, but with a larger uncertainty. The pull of $G_F^\text{EW}$ away from $G_F^\mu$ is mainly driven by $M_W$, $\sin^2\theta_W$ from hadron colliders, $A_\ell$, and $A_{\rm FB}^{0, \ell}$.

\begin{table}[t!]
	\centering
	\begin{tabular}{c c}
	 \toprule
		\begin{tabular}{l r r }	
			$M_W\,[\text{GeV}]$ & ~\cite{Zyla:2020zbs} & $80.379(12)$  \\
			$\Gamma_W\,[\text{GeV}]$ & ~\cite{Zyla:2020zbs} & $2.085(42)$  \\
			$\text{BR}(W\to \text{had})$ & ~\cite{Zyla:2020zbs} & $0.6741(27)$  \\
			$\text{BR}(W\to \text{lep})$ & ~\cite{Zyla:2020zbs} & $0.1086(9)$  \\
			$\text{sin}^2\theta_\text{eff($Q_\text{FB}$)}$ & ~\cite{Zyla:2020zbs}  & $0.2324(12)$  \\
			$\text{sin}^2\theta_{\rm eff(Tevatron)}$ & ~\cite{Aaltonen:2018dxj}  & $0.23148(33)$ \\
			$\text{sin}^2\theta_{\rm eff(LHC)}$ & ~\cite{Aaij:2015lka,Aad:2015uau,ATLAS:2018gqq,Sirunyan:2018swq}  & $0.23129(33)$ \\
			$\Gamma_Z\,[\text{GeV}]$ &~\cite{ALEPH:2005ab} &$2.4952(23)$ \\
			$\sigma_h^{0}\,[\text{nb}]$ &~\cite{ALEPH:2005ab} &$41.541(37)$ \\
			$P_{\tau}^{\rm pol}$ &~\cite{ALEPH:2005ab} &$0.1465(33)$ \\
		\end{tabular} &
		\begin{tabular}{l r r}
			$A_{\ell}$ &~\cite{ALEPH:2005ab} &$0.1513(21)$  \\
			$R^0_{\ell}$ &~\cite{ALEPH:2005ab} &$20.767(25)$ \\
			$A_{\rm FB}^{0, \ell}$&~\cite{ALEPH:2005ab} &$0.0171(10)$   \\
			$R_{b}^{0}$ &~\cite{ALEPH:2005ab} &$0.21629(66)$\\
			$R_{c}^{0}$ &~\cite{ALEPH:2005ab} &$0.1721(30)$ \\
			$A_{\rm FB}^{0,b}$ &~\cite{ALEPH:2005ab} &$0.0992(16)$\\ 
			$A_{\rm FB}^{0,c}$ &~\cite{ALEPH:2005ab} &$0.0707(35)$ \\
			$A_{b}$ &~\cite{ALEPH:2005ab} &$0.923(20)$ \\
			$A_{c}$ &~\cite{ALEPH:2005ab} &$0.670(27)$ \\
		\end{tabular}\\
		\botrule
	\end{tabular}
	\caption{EW observables included in our global fit together with their current experimental values.\label{ObsEW}}
\end{table}

\begin{table}[t!]
	\centering
	\begin{tabular}{l r}
		\hline\hline
		Parameter & Prior \\
		\hline
		$\alpha\times 10^3$~\cite{Zyla:2020zbs} & $7.2973525664(17)$ \\
		$\Delta\alpha_\text{had}\times 10^4$~\cite{Davier:2019can,Keshavarzi:2019abf} & $276.1(1.1)$ \\
		$\alpha_s(M_Z)$~\cite{Zyla:2020zbs,Aoki:2019cca} & $0.1179(10)$\\
		$M_Z\,\,[{\rm GeV}]$~\cite{Zyla:2020zbs,Barate:1999ce,Abbiendi:2000hu,Abreu:2000mh,Acciarri:2000ai} & $91.1876(21)$\\
		$M_H\,\,[{\rm GeV}]$~\cite{Zyla:2020zbs,Aad:2015zhl,Aaboud:2018wps,Sirunyan:2017exp} & $125.10(14)$ \\
		$m_{t}\,\,[{\rm GeV}]$~\cite{Zyla:2020zbs,Khachatryan:2015hba,TevatronElectroweakWorkingGroup:2016lid,Aaboud:2018zbu,Sirunyan:2018mlv}& $172.76(30)$ \\
		\hline\hline
	\end{tabular}
	\caption{Parameters of the EW fit together with their (Gaussian) priors. \label{ParamEW}}
\end{table}

As a third possibility, one can determine the Fermi constant from superallowed $\beta$ decays, taking $V_{us}$ from kaon or $\tau$ decays and assuming CKM unitarity ($|V_{ub}|$ is also needed, but the impact of its uncertainty is negligible). This is particularly interesting given recent hints for the apparent violation of first-row CKM unitarity, known as the Cabibbo angle anomaly (CAA). The significance of the tension crucially depends on the radiative corrections applied to $\beta$ decays~\cite{Marciano:2005ec,Seng:2018yzq,Seng:2018qru,Gorchtein:2018fxl,Czarnecki:2019mwq,Seng:2020wjq,Hayen:2020cxh,Hardy:2020qwl}, but also on the treatment of tensions between $K_{\ell 2}$ and $K_{\ell 3}$ 
decays~\cite{Moulson:2017ive} and the constraints from $\tau$ decays~\cite{Amhis:2019ckw}, see Ref.~\cite{Crivellin:2020lzu} for more details. 
In the end, quoting a significance around $3\sigma$ should give a realistic representation of the current situation, and for definiteness we will thus use the estimate of first-row CKM unitarity from Ref.~\cite{Zyla:2020zbs} 
\begin{align}
\big|V_{ud}\big|^2+\big|V_{us}\big|^2+\big|V_{ub}\big|^2
 = 0.9985(5).
\label{1throw}
\end{align}
In addition, we remark that there is also a deficit in the first-column CKM unitarity relation~\cite{Zyla:2020zbs}
\begin{equation}
\big|V_{ud}\big|^2+\big|V_{cd}\big|^2+\big|V_{td}\big|^2 = 0.9970(18),
\end{equation}
less significant than Eq.~\eqref{1throw}, but suggesting that if the deficits were due to BSM effects, they would likely be related to $\beta$ decays.  For the numerical analysis, we will continue to use Eq.~\eqref{1throw} given the higher precision.  
The deficit in Eq.~\eqref{1throw} translates to
\begin{equation}
\label{GFCKM}
G_F^\text{CKM} = 0.99925(25)\times G_F^\mu =1.16550(29)\times 10^{-5}\GeV^{-2}.
\end{equation}

Comparing the three independent determinations of $G_F$ in Fig.~\ref{GFplot}, one finds the situation in which $G_F^\text{EW}$ lies above $G_F^\mu$ by $2\sigma$,  $G_F^\text{CKM}$ below $G_F^\mu$ by $3\sigma$, and the tension between $G_F^\text{EW}$ and $G_F^\text{CKM}$ amounts to $3.4\sigma$. To bring all three determinations into agreement within $1\sigma$,  an effect in at least two of the underlying processes is thus necessary. This leads us to study BSM contributions to
\begin{enumerate}
	\vspace{-1mm}
	\item $\mu\to e\nu\nu$ transitions,
		\vspace{-1mm}
	\item $Z\to \ell\ell,\nu\nu$, $\alpha_2/\alpha$, $M_Z/M_W$,
		\vspace{-1mm}
	\item superallowed $\beta$ decays, 
\end{enumerate}
	\vspace{-1mm}
where the second point gives the main observables in the EW fit, with $\alpha_2/\alpha$ a proxy for the ratio of $SU(2)_L$ and $U(1)_Y$ couplings. We do not consider the possibility of BSM effects in kaon, $\tau$, or $D$ decays, as this would require a correlated effect with a relating symmetry. Furthermore, as shown in Ref.~\cite{Crivellin:2020lzu}, the sensitivity to a BSM effect in superallowed $\beta$ decays is enhanced by a factor $|V_{ud}|^2/|V_{us}|^2$ compared to kaon, $\tau$, or $D$ decays. This can also be seen from Eq.~\eqref{1throw} as $|V_{ud}|$ gives the dominant contribution.

BSM explanations of the discrepancies between these determinations of $G_F$ have been studied in the literature in the context of the CAA~\cite{Belfatto:2019swo,Bryman:2019bjg,Coutinho:2019aiy,Crivellin:2020lzu,Capdevila:2020rrl,Endo:2020tkb,Crivellin:2020ebi,Kirk:2020wdk,Alok:2020jod,Crivellin:2020oup,Crivellin:2020klg,Crivellin:2021egp}. In this Letter, we will analyze possible BSM effects  
in all three $G_F$ determinations using an EFT approach with gauge-invariant dimension 6 operators~\cite{Buchmuller:1985jz,Grzadkowski:2010es}.

\section{SMEFT analysis}

Dimension-$6$ operators that can explain the differences among the  determinations of $G_F$ can be grouped into the following classes
\begin{enumerate}[A.]
		\vspace{-1mm}
	\item four-fermion operators in $\mu\to e\nu\nu$,
		\vspace{-1mm}
	\item four-fermion operators in $u\to d e\nu$,
		\vspace{-1mm}
	\item modified $W$--$u$--$d$ couplings,
		\vspace{-1mm}
	\item modified $W$--$\ell$--$\nu$ couplings,
		\vspace{-1mm}
    \item other operators affecting the EW fit.
    	\vspace{-1mm}
\end{enumerate}
Global fits to a similar set of effective operators have been considered in Refs.~\cite{Han:2004az,Cirigliano:2009wk,Falkowski:2014tna,Falkowski:2015krw,Ellis:2018gqa,Skiba:2020msb}, here, we will concentrate directly on the impact on $G_F$ determinations, following the conventions of Ref.~\cite{Grzadkowski:2010es}.

\subsection{Four-fermion operators in $\boldsymbol{\mu\to e\nu\nu}$}

Not counting flavor indices, there are only two operators 
that can generate a neutral current involving four leptons:
\begin{align}
Q_{\ell \ell }^{ijkl} = {{\bar \ell }_i}{\gamma ^\mu }{\ell _j}{{\bar \ell }_k}{\gamma ^\mu }{\ell _l},\qquad
Q_{\ell e}^{ijkl} &= {{\bar \ell }_i}{\gamma ^\mu }{\ell _j}{{\bar e}_k}{\gamma ^\mu }{e_l}.
\end{align}
Not all flavor combinations are independent, e.g., 
$Q_{\ell \ell }^{ijkl}=Q_{\ell \ell }^{klij}=Q_{\ell \ell }^{ilkj}=Q_{\ell \ell }^{kjil}$ due to Fierz identities and $Q_{\ell \ell(e) }^{jilk}=Q_{\ell \ell(e) }^{ijkl*}$ due to Hermiticity. 
Instead of summing over flavor indices, it is easiest to absorb these terms into a redefinition of the operators whose latter two indices are $12$, which contribute directly to $\mu\to e\nu\nu$. Therefore, we have to consider 9 different flavor combinations for both operators:

{\setlength{\leftmargini}{12pt}
\begin{enumerate}
		\vspace{-1mm}
 \item $Q_{\ell \ell }^{2112}$
contributes to the SM amplitude (its coefficient is real by Fierz identities and Hermiticity). Therefore, it can give a constructive or destructive effect in the muon lifetime and does not affect the Michel parameters~\cite{Michel:1949qe,Michel:1954eua,Kinoshita:1957zz,Scheck:1977yg,Fetscher:1986uj,Danneberg:2005xv,MacDonald:2008xf,Bayes:2011zza}. In order to bring $G_F^\text{CKM}$ and $G_F^\mu$ into agreement at $1\sigma$ we need
	\begin{equation}
	C_{\ell \ell }^{2112}\approx -1.4\times 10^{-3} G_F.
	\end{equation} 
	This Wilson coefficient is constrained by LEP searches for $e^+e^-\to \mu^+\mu^-$~\cite{Schael:2013ita}
	\begin{equation}
	 - \frac{4\pi}{(9.8\TeV)^2} < C_{\ell \ell }^{1221} < \frac{4\pi}{(12.2\TeV)^2},
	 \end{equation}
	 a factor $8$ weaker than preferred by the CAA, but within reach of future $e^+e^-$ colliders.

	\vspace{-1mm}
\item Even though $Q_{\ell e}^{2112}$ has a vectorial Dirac structure, it leads to a scalar amplitude after applying Fierz identities. Its interference with the SM amplitude is usually expressed in terms of the Michel parameter $\eta=\Re C_{\ell e }^{2112}/(2\sqrt{2}G_F)$, leading to a correction $1-2\eta m_e/m_\mu$. In the absence of right-handed neutrinos the restricted analysis from Ref.~\cite{Danneberg:2005xv} applies, constraining the shift in $G_F^\mu$ to $0.68\times 10^{-4}$, well below the required effect to obtain $1\sigma$ agreement with $G_F^\text{CKM}$ or $G_F^\text{EW}$.  
	\vspace{-1mm}	
\item The operators $Q_{\ell \ell (e) }^{1212}$ could contribute to muon decay as long as the neutrino flavors are not detected. To reconcile $G_F^\text{CKM}$ and $G_F^\mu$ within $1\sigma$ we need $|C_{\ell \ell }^{1212}|\approx 0.045\,G_F$ or 	$|C_{\ell e }^{1212}|\approx 0.09\,G_F$. Both solutions are excluded by muonium--anti-muonium oscillations ($M=\mu^+ e^-$)~\cite{Willmann:1998gd}
	\begin{equation}
	{\cal P}(\bar{M}\text{--}M)<8.3\times 10^{-11}/S_B,
	\label{exp:mmbar}
	\end{equation}
	with correction factor $S_B=0.35$ ($C_{\ell \ell }^{1212}$) and $S_B=0.78$ ($C_{\ell e }^{1212}$) for the extrapolation to zero magnetic field. 
	Comparing to the prediction for the rate~\cite{Feinberg:1961zza,Hou:1995np,Horikawa:1995ae}
	\begin{align}
	{\cal P}(\bar{M}\text{--}M) &=
	\frac{8(\alpha \mu_{\mu e})^6\tau_\mu^2 G_F^2}{\pi^2}
	\, \left|{C_{\ell \ell (e) }^{1212}}/{G_F}\right|^2\notag\\
	&=3.21\times 10^{-6} \left|{C_{\ell \ell (e) }^{1212}}/{G_F}\right|^2,
	\end{align}
	with reduced mass $\mu_{\mu e}=m_\mu m_e/(m_\mu+m_e)$, the limits become
	$|C_{\ell \ell (e) }^{1212}|<8.6 (5.8)\times 10^{-3}G_F$.
			
	\vspace{-1mm}		
\item For $Q_{\ell \ell (e) }^{1112}$ again numerical values of $
|C_{\ell \ell (e) }^{1112}|\approx 0.09\, G_F$ are 
preferred (as for all the remaining Wilson coefficients in this list). Both operators give tree-level effects in $\mu\to 3 e$, e.g.,
	\begin{equation}
	 \text{Br}\left[ {\mu  \to 3e} \right] = \frac{{m_\mu ^5\tau_\mu}}{{768{\pi ^3}}}{\left| {C_{\ell \ell }^{1112}} \right|^2}=0.25\bigg|\frac{C_{\ell \ell }^{1112}}{G_F}\bigg|^2,
	\end{equation}
	which exceeds the experimental limit on the branching ratio of $1.0\times 10^{-12}$~\cite{Bellgardt:1987du} by many orders of magnitude (the result for $C_{\ell e }^{1112}$ is smaller by a factor $1/2$).
	\vspace{-1mm}		
\item The operators $Q_{\ell \ell (e) }^{2212}$ and $Q_{\ell \ell (e) }^{3312}$ contribute at the one-loop level to $\mu\to e$ conversion and $\mu\to 3e$ and at the two-loop level to $\mu\to e\gamma$~\cite{Crivellin:2017rmk}. Here the current best bounds come from $\mu\to e$ conversion. Using Table~3 in Ref.~\cite{Crivellin:2017rmk} we have
\begin{align}
 \left| {C_{\ell \ell }^{3312}} \right| &<6.4\times 10^{-5} G_F,\notag\\
 \left| {C_{\ell \ell }^{2212}} \right| &<2.8\times 10^{-5} G_F,
\end{align}
excluding again a sizable BSM effect, and similarly for $Q_{\ell e }^{3312}$ and $Q_{\ell e }^{2212}$.
	
	\vspace{-1mm}
\item $Q_{\ell \ell (e) }^{2312}$, $Q_{\ell \ell (e) }^{3212}$ and $Q_{\ell \ell (e) }^{1312}$, $Q_{\ell \ell (e) }^{3112}$  contribute to $\tau\to \mu \mu e$ and $\tau\to \mu ee$, respectively, 
which excludes a sizable effect in analogy to $\mu\to 3e$ 
above~\cite{Hayasaka:2010np,Lees:2010ez,Amhis:2019ckw}. 
	\vspace{-1mm}
\end{enumerate}	}
Other four-quark operators can only contribute via loop effects, which leads us to conclude that the only viable mechanism proceeds via a modification of the SM operator $Q_{\ell \ell }^{2112}$.

\subsection{Four-fermion operators in $\boldsymbol{d\to u e\nu}$}

First of all, the operators $Q_{\ell equ}^{\left( 1 \right)1111}$ and $Q_{\ell equ}^{\left( 3 \right)1111}$ give rise to $d\to u e\nu$ scalar amplitudes. Such amplitudes lead to enhanced effects in $\pi\to\mu\nu/\pi\to e\nu$ with respect to $\beta$ decays and therefore can only have a negligible impact on the latter once the stringent experimental bounds~\cite{Aguilar-Arevalo:2015cdf,Zyla:2020zbs} are taken into account. Furthermore, the tensor amplitude generated by $Q_{\ell equ}^{\left( 3 \right)ijkl}$ has a suppressed matrix element in $\beta$ decays.

Therefore, we are left with $Q_{\ell q}^{\left( 3 \right)1111}$, for which we only consider the flavor combination that leads to interference with the SM. The CAA prefers
$C_{\ell q}^{\left( 3 \right)1111}\approx 0.7\times 10^{-3}G_F$.
Via $SU(2)_L$ invariance, this operator generates effects in neutral-current (NC) interactions
\begin{equation}
{{\cal L}_{{\rm{NC}}}} = C_{\ell q}^{\left( 3 \right)1111}\left( {\bar d{\gamma ^\mu }{P_L}d - \bar u{\gamma ^\mu }{P_L}u} \right)\bar e{\gamma _\mu }{P_L}e.
\end{equation}
Note that since the SM amplitude for $\bar uu (\bar dd)\to e^+e^-$, at high energies, has negative (positive) sign, we have constructive interference in both amplitudes. Therefore, the latest nonresonant dilepton searches by ATLAS~\cite{Aad:2020otl} naively lead to 
\begin{equation}
C_{\ell q}^{\left( 3 \right)1111}	\lesssim 1.6\times10^{-3}G_F,
\end{equation} 
which implies that bringing $G_F^\text{CKM}$ into $1\sigma$ agreement with $G_F^\mu$
via four-fermion operators affecting $d\to u e\nu$ transitions is still possible. However, ATLAS derived the bound for the $SU(2)_L$ singlet operator, which means that the actual constraint for triplet operators is stronger, as it leads to constructive interference in both the up- and the down-quark channels.
In consequence, the required value of
$C_{\ell q}^{\left( 3 \right)1111}$ lies at the border of the ATLAS constraint.

\subsection{Modified $\boldsymbol{W}$--$\boldsymbol{u}$--$\boldsymbol{d}$ couplings}

There are only two operators modifying the $W$ couplings to quarks
\begin{align}
Q_{\phi q}^{\left( 3 \right)ij} &= {\phi ^\dag }i
\overset{\leftrightarrow}{D}^I_\mu
	\phi {{\bar q}_i}{\gamma ^\mu }{\tau ^I}{q_j},\notag\\
Q_{\phi ud}^{ij} &= {\tilde \phi ^\dag }iD_\mu \phi {{\bar u}_i}{\gamma ^\mu }{d_j}.
\end{align}
First of all, $Q_{\phi ud}^{ij}$ generates right-handed $W$--quark couplings~\cite{Bernard:2007cf,Crivellin:2009sd,Buras:2010pz,Crivellin:2014zpa,Alioli:2017ces}, which can solve the CAA. In addition, a right-handed $W$--$u$--$s$ coupling could also account for the difference between $K_{\ell2}$ and $K_{\ell3}$ decays~\cite{Grossman:2019bzp}. $Q_{\phi q}^{\left( 3 \right)ij}$ generates modifications of the left-handed $W$--quark couplings and data prefer
\begin{equation}
C_{\phi q}^{\left( 3 \right)11} \approx -0.7\times 10^{-3}G_F.
\end{equation}
Due to $SU(2)_L$ invariance, in general effects in $D^0$--$\bar D^0$ and $K^0$--$\bar K^0$ mixing as well as in $\Gamma[Z\to {\rm hadrons}]/\Gamma[Z\to {\rm leptons}]$ are generated. However, the former bounds can be avoided by a $U(2)$ flavor symmetry and the latter by simultaneous contributions to $Q_{\phi q}^{\left( 1 \right)ij}$. For a detailed analysis we refer the reader to Ref.~\cite{Belfatto:2021jhf}.

\subsection{Modified $\boldsymbol{W}$--$\boldsymbol{\ell}$--$\boldsymbol{\nu}$ couplings}

Only the operator
\begin{align}
 Q_{\phi \ell }^{(3)ij}={\phi ^\dag }i
\overset{\leftrightarrow}{D}^I_\mu
	\phi {{\bar \ell}_i}{\gamma ^\mu }{\tau ^I}{\ell_j}
\end{align}
generates modified $W$--$\ell$--$\nu$ couplings at tree level. In order to avoid the stringent bounds from charged lepton flavor violation, the off-diagonal Wilson coefficients, in particular $C_{\phi \ell }^{(3)12}$, must be very small. Since they also do not generate amplitudes interfering with the SM ones, their effect can be neglected. While $C_{\phi \ell }^{(3)11}$ affects $G_F^\mu$ and $G_F^\text{CKM}$ in the same way, $C_{\phi \ell }^{(3)22}$ only enters in muon decay. Therefore, agreement between $G_F^\mu$ and $G_F^\text{CKM}$ can be achieved by $C_{\phi \ell }^{(3)11}<0$, $C_{\phi \ell }^{(3)22}>0$, and $|C_{\phi \ell }^{(3)22}|<|C_{\phi \ell }^{(3)11}|$ without violating lepton flavor universality tests such as $\pi(K)\to\mu\nu/\pi(K)\to e\nu$ or $\tau\to\mu\nu\nu/\tau(\mu)\to e\nu\nu$~\cite{Coutinho:2019aiy,Crivellin:2020lzu,Pich:2013lsa}. However, $C_{\phi \ell }^{(3)ij}$ also affects $Z$ couplings to leptons and neutrinos, which enter the global EW fit.

\subsection{Electroweak fit}

\begin{figure}[t]
	\includegraphics[width=0.9\linewidth]{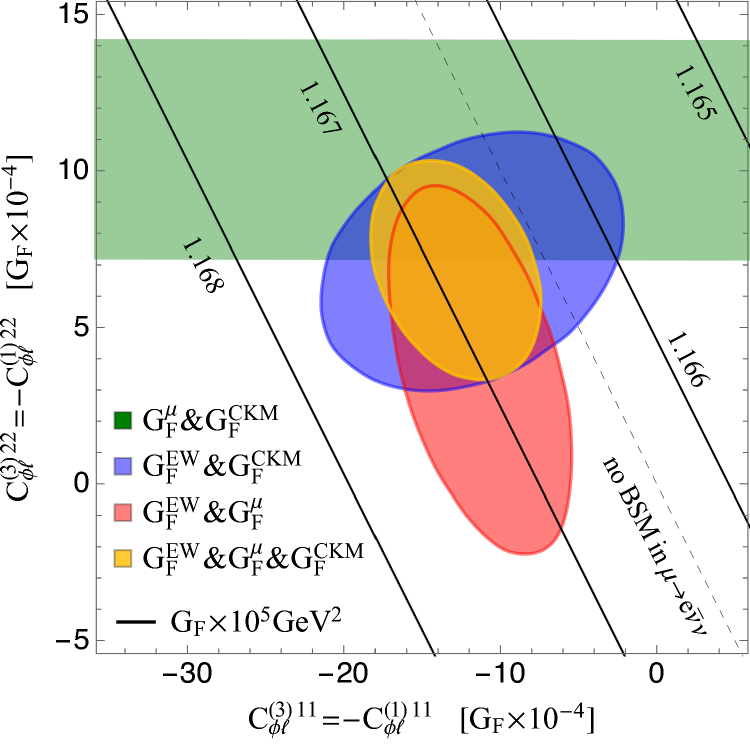}
	\caption{Example of the complementarity between the $G_F$ determinations from muon decay ($G_F^\mu$), CKM unitarity ($G_F^\text{CKM}$), and the global EW fit ($G_F^\text{EM}$) in case of $C_{\phi \ell }^{(3)ii}=-C_{\phi \ell }^{(1)ii}$, corresponding to modifications of neutrino couplings to gauge bosons (the EW fit also includes $\tau\to\mu\nu\nu/\tau(\mu)\to e\nu\nu$~\cite{Amhis:2019ckw,Zyla:2020zbs,Pich:2013lsa}). Here, we show the preferred $1\sigma$ regions obtained by requiring that two or all three $G_F$ determinations agree. The contour lines show the value of the Fermi constant extracted from muon decay once BSM effects are taken into account.} \label{FinalPlot}
\end{figure}

The impact of modified gauge-boson--lepton couplings on the global EW fit, generated by $Q_{\phi \ell}^{\left( 3 \right)ij}$ and 
\begin{align}
Q_{\phi \ell}^{\left( 1 \right)ij} &= {\phi ^\dag }i\overset{\leftrightarrow}{D}_\mu \phi {{\bar \ell}_i}{\gamma ^\mu }{\ell_j},
\end{align}
can be minimized by only affecting $Z\nu\nu$ but not $Z\ell\ell$, by imposing $C_{\phi \ell }^{(1)ij}=-C_{\phi \ell }^{(3)ij}$. In this way, in addition to the Fermi constant, only the $Z$ width to neutrinos changes and the fit improves significantly compared to the SM~\cite{Coutinho:2019aiy}, see Fig.~\ref{FinalPlot} for the preferred parameter space. One can even further improve the fit by assuming $C_{\phi \ell }^{(1)11}=-C_{\phi \ell }^{(3)11}$, $C_{\phi \ell }^{(1)22}=-3C_{\phi \ell }^{(3)22}$, which leads to a better description of $Z\to\mu\mu$ data~\cite{Crivellin:2020ebi,Kirk:2020wdk}. Furthermore, the part of the tension between $G_F^\text{EW}$ and $G_F^{\mu}$ driven by the $W$ mass can be alleviated by the operator $Q_{\phi W B}=\phi^\dagger \tau^I\phi W^I_{\mu\nu}B^{\mu\nu}$. 

\section{Conclusions and outlook}

Since the Fermi constant is determined extremely precisely by the muon lifetime, Eq.~\eqref{Mulan}, constraining BSM effects via $G_F$
is limited entirely by the precision of the second-best determination. In this Letter we derived in a first step two alternative independent determinations, from the EW fit, Eq.~\eqref{GFEW}, and superallowed $\beta$ decays using CKM unitarity, Eq.~\eqref{GFCKM}. The latter determination 
is more precise than the one from the EW fit, even though the precision of $G_F^\text{EW}$ increased by a factor 5 compared to Ref.~\cite{Marciano:1999ih}. Furthermore, as shown in Fig.~\ref{GFplot}, the two determinations display a tension of $2\sigma$ and $3\sigma$ compared to $G_F^\mu$, respectively. 

In a second step, we investigated how these hints of BSM physics can be explained within SMEFT. For BSM physics in $G_F^\mu$ we were able to rule out all four-fermion operators, except for $Q_{\ell\ell}^{2112}$, which generates a SM-like amplitude, and modified $W$--$\ell$--$\nu$ couplings, from $Q_{\phi \ell }^{(3)ij}$. Therefore, both constructive and destructive interferences are possible, which would bring $G_F^\mu$ into agreement with $G_F^\text{CKM}$ or $G_F^\text{EW}$, respectively, at the expense of increasing the tension with the other determination. To achieve a better agreement among the three different values of $G_F$, also BSM effects in $G_F^\text{CKM}$ and/or $G_F^\text{EW}$ are necessary. In the case of $G_F^\text{CKM}$ only a single four-fermion operator, $Q_{\ell q}^{\left( 3 \right)1111}$, and $Q_{\phi \ell }^{(3)ij}$, $Q_{\phi q}^{(3)ij}$ remain.  Finally, modified gauge-boson--lepton couplings, via $Q_{\phi \ell}^{\left( 3 \right)ij}$ and $Q_{\phi \ell}^{\left( 1 \right)ij}$, can not only change $G_F^\text{CKM}$ and $G_F^{\mu}$, but also affect the EW fit via the $Z$-pole observables, which can further improve the global agreement, see Fig.~\ref{FinalPlot}. 
This figure also demonstrates the advantage of interpreting the tensions in terms of $G_F$, defining a transparent benchmark for comparison both in SMEFT and concrete BSM scenarios, and allows one to constrain the amount of BSM contributions to muon decay.  

Our study highlights the importance of improving the precision of the alternative independent determinations of $G_F^\text{CKM}$ and $G_F^\text{EW}$ in order to confirm or refute BSM contributions to the Fermi constant. Concerning $G_F^\text{CKM}$, improvements in the determination of $|V_{ud}|$ should arise from advances in nuclear-structure~\cite{Cirgiliano:2019nyn,Martin:2021bud} and EW radiative corrections~\cite{Feng:2020zdc}, while experimental developments~\cite{Fry:2018kvq,Soldner:2018ycf,Wang:2019pts,Serebrov:2019puy,Gaisbauer:2016kan,Ezhov:2018cta,Callahan:2018iud} could make the determination from neutron decay~\cite{Pattie:2017vsj,Markisch:2018ndu,Czarnecki:2018okw} competitive and, in combination with $K_{\ell 3}$ decays, add another complementary
constraint on $|V_{ud}|/|V_{us}|$ via pion $\beta$ decay~\cite{Pocanic:2003pf,Czarnecki:2019iwz}. 
Further, improved measurements of 
 $|V_{cd}|$ from $D$ decays~\cite{Ablikim:2019hff} could bring the precision of the first-column CKM unitarity relation close to the first-row one, which, in turn, could be corroborated via improved
 $|V_{us}|$ determinations from $K_{\ell3}$ decays~\cite{Yushchenko:2017fzv,Junior:2018odx,Babusci:2019gqu}. 
The precision of $G_F^\text{EW}$ will profit in the near future from LHC measurements of $m_t$ and $M_W$, in the mid-term from the Belle-II EW precision program~\cite{Kou:2018nap}, and in the long-term from future $e^+e^-$ colliders such as the FCC-ee~\cite{Abada:2019zxq}, ILC~\cite{Baer:2013cma}, CEPC~\cite{An:2018dwb}, or CLIC~\cite{Aicheler:2012bya}, which could achieve a precision at the level of $10^{-5}$.

\begin{acknowledgments}
We thank David Hertzog and Klaus Kirch for valuable discussions, 
and the ATLAS collaboration, in particular Noam Tal Hod, for clarifications concerning the analysis of Ref.~\cite{Aad:2020otl}.
Support by the Swiss National Science Foundation, under Project Nos.\ PP00P21\_76884 (A.C., C.A.M.) and PCEFP2\_181117 (M.H.) is gratefully acknowledged.
\end{acknowledgments}

\bibliography{GF}

\end{document}